\documentstyle[epsf,12pt]{article}

\begin{document}
\epsfysize3cm
\epsfbox{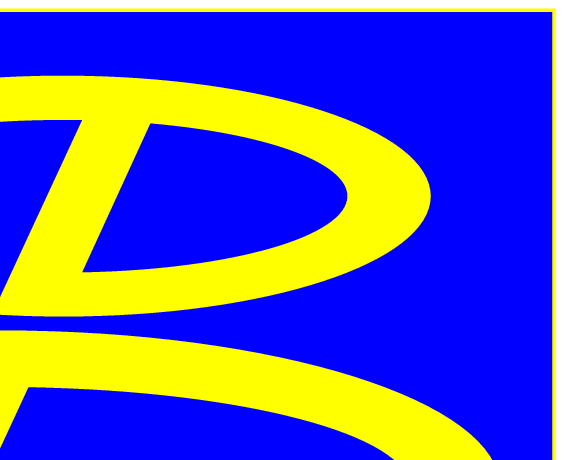}
\vskip -3cm
\noindent
\hspace*{4.5in}NTU-HEP 96-1\\
\hspace*{4.5in}NKNU-HEP 96-1\\
\hspace*{4.5in}BELLE Preprint 96-1 \\
\hspace*{4.5in}KEK Preprint 96-4 \\
\hspace*{4.5in}NLCTC-HEP 96-1
\begin{center}
\vskip 2cm
{\Large  \bf
Measurement of Radiation Damage 
on Silica Aerogel \v Cerenkov Radiator\footnote{
submitted to NIM-A.}
}\\
\vskip 1cm
S.K.~Sahu$^a$\footnote{Internet address: sahu@kekvax.kek.jp}, 
M.Z.~Wang$^b$, R. Suda$^c$, R. Enomoto$^d$,
K.C.~Peng$^a$,
C.H.~Wang$^e$, 
I.~Adachi$^d$, 
M.~Amami$^f$,
Y.H.~Chang$^g$,
R.S.~Guo$^b$,
K.~Hayashi$^d$, 
T.~Iijima$^d$,
T.~Sumiyoshi$^d$, 
Y.~Yoshida$^h$\\ 
\vskip 1cm
{\it
$^a$National Taiwan University, Taipei, Taiwan\\
$^b$National Kaohsiung Normal University, Kaohsiung, Taiwan\\
$^c$Tokyo Metropolitan University, Tokyo 192-03\\
$^d$National Laboratory for High Energy Physics, KEK, Ibaraki 305\\
$^e$National Lien Ho College of Tech. and Commerce, Miao Li, Taiwan\\
$^f$Saga University, Saga 840\\
$^g$National Central University, Chung-Li, Taiwan\\
$^h$Toho University, Chiba 274\\
}
\end{center}
\newpage
\begin{abstract}
We measured the radiation damage on silica aerogel \v Cerenkov radiators
originally developed for the $B$-factory experiment at KEK.
Refractive index of the aerogel samples ranged from 1.012 to 1.028. 
The samples were irradiated up to
9.8~MRad of equivalent dose. Measurements of transmittance and refractive index
were carried out and these samples were found to be radiation hard.
Deteriorations in transparency and changes of refractive index were 
observed to be less than
1.3\% and 0.001 at 90\% confidence level, respectively.
Prospects of using aerogels under high-radiation environment
are discussed.
\end{abstract}
\newpage
\section{Introduction}
Silica aerogels(aerogels) are a colloidal form of glass, in which globules of 
silica are connected in three dimensional networks  with siloxan
bonds.
They are solid,
very light, transparent and their refractive index can be controlled in 
the production process.  Many high energy and nuclear physics experiments 
have used aerogels instead of  pressurized gas for 
their \v Cerenkov counters
\cite{used1,used2,used3,used4,used5,used6,used7,used8,used9,used10}.
 
Several experiments in near future operating under very 
high dose of radiation are likely to employ aerogel
\v Cerenkov counters for 
their particle identification systems\cite{aerogel,slactdr,loi,tdr}. 
Stability of aerogels under such circumstances has, however, not been 
studied yet. In this report we for the first time show the effect
of radiation damage on aerogel samples produced
by some of us at KEK\cite{aerogel,loi,tdr}. 
 
In the next section, we quantify the \v Cerenkov light production
in aerogels. Prospects of using aerogels in present
and future experiments are explored in the third section. Experimental
setup and results are described in fourth and fifth sections,
respectively.  Conclusion is stated in the 
sixth section of this report.
\section{Silica Aerogels for KEK B-Factory}
 
The production method of aerogels used for the KEK B-factory experiment
can be found in Ref. \cite{aerogel,loi,tdr}. 
Accessible range of refractive index
($n$) in this method is between 1.006 and 1.060, and it can be controlled
to a level of $\delta n$=0.0004. 
Typical size of a tile was
12cm $\times$ 12cm $\times$ 2cm within a tolerance of 0.3\%.
Absorption and scattering lengths as functions of incident 
wavelength were obtained
from a transmittance measurement with a
spectrophotometer\cite{monochrometer} using the following relations.
We defined a transmission length $\Lambda$ as;
\begin{equation}
T/T_0 = exp \left(-t/\Lambda \right),
\end{equation}
where T/T$_{0}$ is transmittance, and $t$ is thckness of aerogels.
Then we found the absorption and scattering length by fitting $\Lambda$
with the following equations;
\begin{equation}
\frac{1}{\Lambda} = 
\frac{1}{\Lambda_{abs}} + \frac{1}{\Lambda_{scat}}
\label{lambda1}
\end{equation}
\begin{equation}
\Lambda_{abs}=a\lambda^2,~\Lambda_{scat}=b\lambda^4,
\label{lambda2}
\end{equation}
where $\Lambda_{abs}$ is absorption length, $\Lambda_{scat}$ is scattering
length, and a, b are free parameters.
The results are shown in 
Figure \ref{transmittance}.
At 400 nm, where a typical photo-multiplier tube (PMT) with a bialkali
photocathode has maximum sensitivity, the absorption 
and scattering lengths were measured to be
$22\pm2$ and $5.2\pm0.5$ cm, respectively.

The \v Cerenkov light yield of our aerogel samples was measured 
using a 3.5 GeV/c $\pi^-$ beam at KEK PS ($\pi 2$ beam line). 
Light yield from 12cm-thick
aerogel ({\it i.e.}, 6 layers) 
was measured by two 3-inch PMTs which were directly attached
to both sides of aerogel surfaces.
The counter size was 12cm $\times$ 12 cm $\times$ 12 cm .
 Surfaces other than the photocathode
area were covered by GORETEX white reflector\cite{gortex}.
Measurements of $N_{pe}$ and $N_0$, as defined below, 
are given in Table \ref{n0}. Number of photoelectrons
($N_{pe}$) is given by Frank-Tamm's equation\cite{franc_tamm}~:
\begin{equation}
{dN_{pe} \over dE} = \left( {\alpha \over \hbar c} \right)
\cdot L \cdot z^2 \cdot \sin ^2 \theta _C \cdot \epsilon _{QE} (E)
\cdot \epsilon ,
\label{frank-tamm}
\end{equation}
where $\alpha$ is the fine structure constant,
$L$ thickness of radiator, $z$ charge of an incident particle, 
$\theta _{C}$
\v Cerenkov angle, $\epsilon _{QE}(E)$ quantum efficiency of the PMTs,
and $\epsilon$ detection efficiency including light absorption and light
collection. 
Energy-integrated quantum efficiency of PMTs used in this experiment was
measured to be
\begin{equation}
\int \epsilon _{QE} (E) dE = 0.46 eV.
\label{qe}
\end{equation}
 
Using Equations \ref{frank-tamm} and \ref{qe},
$\epsilon$ was estimated to be $\sim$ 0.6, which is higher than that for
any other existing aerogel counters. 
The \v Cerenkov Quality Factor $N_0$ defined as;
\begin{equation}
N_0 =  \left( {\alpha \over \hbar c} \right)
\cdot \int \epsilon _{QE} 
\cdot \epsilon \cdot dE,
\label{n0def}
\end{equation}
for samples of different refractive
indices are given in Table \ref{n0}.
For our aerogels,
$N_0$ was typically $\sim$100cm$^{-1}$, independent of the refractive
index in the range of 1.01 $< n <$ 1.03.
\section{Use of Aerogels in High-Radiation Environment}
\subsection{B-Factory}
$B$-factories are high luminosity colliders where copious amount of
$B$-mesons are produced and their decays are studied \cite{slactdr,loi,tdr}. 
The primary goal of these 
machines is to study the CP violation in the heavy quark sector
\cite{km,sanda}.
Because of high luminosity of these machines, corresponding
detectors are subjected to a substantial amount of radiation from 
both physics and background events. 
Particle identification is a vital part of the detectors in these
machines. Threshold aerogel \v Cerenkov counters are being proposed
for $\pi/K/p$ separation in these detectors.
 
\subsubsection{$e^+e^-$ Collider}
Two asymmetric $B$-factories, producing $B$-mesons by annihilating 
electrons and positrons are being constructed at KEK \cite{loi,tdr} and 
SLAC
\cite{slactdr}. Threshold aerogel \v Cerenkov counters 
are to be used as parts of particle identification system.
Radiation dose in these detectors is estimated to be typically 1 kRad/year
at places closest to the beam-pipe \cite{tdr}.
\subsubsection{Hadron Machine}
An imaging \v Cerenkov counter using aerogel radiator has been proposed
\cite{spot}, which could be a potential candidate for 
a particle identification system at low angle in a generic 
hadron $B$-factories, such as the HERA-$B$ experiment\cite{herab}.
Considering the scattering length of our aerogel
(Figure \ref{transmittance}), aerogel as thin as 1 cm
can be used as a radiator to make a ring image of 
the \v Cerenkov radiation. 
In case of the HERA-$B$ experiment, the radiation dose is
expected to be 10Mrad/year at the innermost edge.
\subsection{Nuclear Science}
Possibility of separating high energy isotopes
by aerogel \v Cerenkov counters is discussed here.
Defining
\begin{equation}
n_0=n-1 \mbox{ and } \beta_0 = 1 - \beta,
\label{definition}
\end{equation}
we obtain 
\begin{equation}
N_{pe} \simeq 2N_0 L z^2 (n_0 - \beta_0).
\end{equation}
using equations \ref{frank-tamm}, \ref{n0def}, and \ref{definition}.
 
Assuming a Poisson distribution for $N_{pe}$, the measurement error
of $\beta$ is calculated to be
\begin{equation}
\delta \beta_0 \simeq {1 \over z} \sqrt{n_0-\beta_0 \over 2N_0 L}.
\end{equation}
The momentum resolution for a nucleon is expressed as
\begin{equation}
{\delta p \over p} \simeq {1 \over z \beta_0} \sqrt{n_0-\beta_0
\over 8 N_0 L}.
\end{equation}
At a little above the threshold, 
{\it i.e.}, $\beta_0 \sim n_0/2$, this becomes
\begin{equation}
{\delta p \over p} \simeq {1 \over z \sqrt{
4 n_0 N_0 L}}.
\end{equation}
Assuming $N_0=100$ cm$^{-1}$, $L$=10 cm, and $n$=1.05, we obtain
a momentum resolution of 0.07/$z$ which is better than 
the typical resolution of magnetic
spectrometers for high-$z$ incident particles {\it i.e.}, heavy nuclei.
Therefore a combination of aerogel counters and a magnetic spectrometer
can separate isotopes. The nuclear charge $z$ can be measured by 
some other devices such as
scintillator and/or combination 
of \v Cerenkov radiators.
Rigidity $R$ can also be measured by the 
magnetic spectrometer.
Therefore we can determine the mass number of the nucleus (A), using the 
relationship
\begin{equation}
Ap = zR.
\end{equation}
For nuclei having the $A/z$ value close to 2, measurement error of A
is calculated to be
\begin{equation}
\delta A \sim 0.14 + x A R,
\end{equation}
where $x$ is the rigidity resolution of the spectrometer  
determined by the position resolution and multiple Coulomb
scattering.
Therefore the limit of separating heavy isotopes is held by
spectrometer resolution. 
For example, with n = 1.05 aerogel and 
for 5 GeV/nucleon region, we can distinguish
nuclei up to
Boron with a  0.5\%-resolution spectrometer and up to Vanadium with 
a 0.1\%-resolution one.
 
In summary, using aerogels, we can distinguish nuclear fragments in heavy ion
collisions at low angle. 
Combination of aerogel counters of several refractive indices can even
replace a magnetic spectrometer as far as the issue of particle 
identification goes.

For the RHIC collider, the radiation dose at a very low angle is considered
to be less than 100krad/year \cite{phenix}.
 
\subsection{Space Experiments}
HEAO-C2 experiment \cite{heao}, which used 
aerogel \v Cerenkov counters, has measured an abundance of heavy nuclei
in high energy cosmic rays. 
Quality of our aerogel is more than four times better
than that used in HEAO-C2, thereby showing that if used, it could 
substantially improve the
quality of the data. In addition, a combination of aerogels and 
a magnetic spectrometer in the experiment could separate 
the isotopes. This would be a completely new method. 
The expected radiation dose is 10 kRad/year at 600 km 
altitude \cite{takahashi}.
 
Recently a space station based experiment -- Anti-Matter Spectrometer
(AMS) has been proposed\cite{ting}, which will look for heavy 
anti-matter nuclei in space. The same technique for isotope separation
discussed above can be used to separate isotopes in this spectrometer.
The AMS will orbit the earth at about 300 km altitude, where
the radiation dose is expected to be several kilorads per year.
 
\section{Experimental Setup}
Optical transparency of the aerogel should be as high as possible
not to lose \v Cerenkov photons inside it. The reflactive index of
the aerogel (n), should be stable during an experiment.
In the present work, two properties of aerogel samples,
transmittance and refractive index were measured 
in order to monitor radiation damage on the samples at the irradiation
facility of National Tsing Hua University (Taiwan).
In the cases of $B$-factories and space experiments, most of the irradiation
is caused by electrons and/or $\gamma$-rays
of critical energies.
We used a Co$^{60}$ $\gamma$-ray source, activity of which was
1320 Curie. 
The error in estimation of radiation dose 
was dominated by the uncertainty in the placement of the sample (0.3cm)
in front of the source
and the ambiguity of 
radius dependent dose. The error at the highest dose value (9.8Mrad of
{\it equivalent~dose} \cite{edose})
was the largest, and it was 17\%.
For each aerogel sample, five or six irradiations
were carried out. 
 
\subsection{Transmittance}
Transmittance of the aerogel samples was measured by observing the 
ratio of photons absorbed in the aerogel volume with and 
without the irradiation.  A schematic diagram of the setup is 
given in Figure~\ref{setupa}. Three aerogel samples of dimensions
12 cm $\times$ 12 cm $\times$ 2 cm  each were stacked together
in a zinc box. This box was kept inside a mother-box having 
an LED light box on one side and a photo-multiplier tube 
on the other. The whole system was enclosed in a big light-tight
box. Inner sides of this box were covered with black clothes to absorb
any stray light.
 
We made two such stacks for each refractive index. One 
was irradiated (RAD-sample) and the other one was kept 
shielded in the irradiation cell (REF-sample). The latter 
is for reference. It is important that both RAD and REF be
kept under same environmental conditions to cancel out any
effects caused by humidity, temperature, dust, or any other
factors other than gamma-ray radiation.  
 
The LED (blue in color) was triggered by an external pulse 
generator to produce bursts of photon. Typically the trigger 
signal had a pulse height of 3.75 V, width of 420 ns, and 
was repeated every 10 msec.  About 400 photo-electrons were produced 
for each burst in the absence of aerogel asmples between the LED 
and the phototube.  The number reduced to about 200 p.e.
when the stack of aerogel was introduced. 
 
We integrated the charge produced by a 2-inch PMT
\cite{r329} for each photon burst 
within a gate of 0.5 $\mu$sec. Ratio of
this integrated charge with aerogel between PMT and LED to 
that without aerogel gives us a measure of the transparency
of that aerogel sample.  We will call this ratio as 
$r_{RAD}$ for irradiated sample and $r_{REF}$ for the 
reference sample. The ratio $r_{abs} = r_{RAD}/r_{REF}$ gives us 
the transparency of irradiated sample with respect to the
reference one.
It may be noted that $r_{abs}$ is sensitive only to radiation damage,
whereas $r_{REF}$ tells us if there is deterioration
such as due to atmospheric conditions. 
After each stage of irradiation, the RAD and REF samples were 
tested for transmittance, and the ratios $r_{REF}$, $r_{RAD}$ and
$r_{abs}$ were calculated.
 
Before the irradiation, $r_{REF}$ and $r_{RAD}$ were measured 
several times to make sure that we get a set of systematically 
consistent consecutive readings.
The dominant systematic error came from the uncertainty of placement of
aerogel crystals. Other minor sources were 
temperature dependence and instabilities
of PMT and LED.  
From these readings we found that the combined error of measurement of 
each transparency ratio, $r_{RAD}$ or $r_{REF}$ was 0.55\%.
Therefore we estimated the error of r$_{abs}$ to be 0.78(=0.55$\sqrt{2}$)\%.
\subsection{Refractive Index}
Refractive index of the aerogel sample was monitored using 
the {\em Prism Formula}
\begin{equation}
n = \frac{\sin(\phi /2 + \alpha /2)}{\sin(\phi /2)},
\label{prism}
\end{equation}
where $n$ is the refractive index, $\phi$ the angle of the prism,
$\alpha$ the angle of the minimum deflection. 
 
The setup for measuring $n$ is shown in Figure~\ref{setupb}.
A red laser (Ar II)
was shone on the corner of the aerogel sample placed
on a rotating table. The aerogel behaved as a prism, and 
deviated the laser beam on a screen 
placed at a distance $l$. By rotating the table 
manually, and checking the minimum distance $d$, angle of the 
minimum deflection
$\alpha$ was found out.  
 
A special sample was prepared which was irradiated in parallel with 
the stacks for transmittance test. Its refractive index was monitored 
by the method described above after each stage of irradiation.
 
We estimated the error of refractive index measurement by calculating
error propagation of the prism formula. It is estimated as ; 
\begin{equation}
\left( {\Delta n \over n} \right) ^2= 
\left[ \left\{ \sin {\alpha \over 2} -
{\cos {\alpha \over 2} \over \tan {\phi \over 2}}\right\}
\cos ^2 \alpha \right]^2
\left\{ {d^2 \over 4l^4} (\Delta l)^2+{1 \over 4 l^2}(\Delta d)^2 \right\}
+{1\over 4} 
{\sin ^2 {\alpha \over 2} \over \sin ^4 {\phi \over 2}} (\Delta \phi)^2.
\label{errorprism}
\end{equation}
For $n$ =1 .012 and 1.028 crystals, 
the accuracy of measuring $l$ was $\Delta l$ = 0.2 cm and that
of $d$ was $\Delta d$ = 0.1 cm.
For $n$ = 1.018 samples, however, the $\Delta d$ was worse (=0.3 cm)
due to the inferior
surface quality, giving rise to a spread of the laser spot on the screen.
Error in measuring the refracting angle of the sample was $\Delta\phi$
 = $0.1^\circ$.
The errors $\Delta n$ were then calculated to be 0.00035 for $n$ = 1.012 and
1.028 samples, and 0.001 for $n$ = 1.018 sample, using 
Equation~\ref{errorprism}.
It may be noted that $\Delta d$ dominates the final error.
 
\section{Results}
\subsection{Transmittance}
Samples of three refractive indices~:  1.012, 1.018 and 1.028 were 
tested. The transparency ratio $r_{abs}$ was measured at several
points, ranging from 1 kRad to 9.8 MRad. The results are
plotted in Figure~\ref{radtra} for each reflactive index.
 
No degradation in transparency is observed in any samples
within experimental errors.
Defining absolute deterioration as the maximum of
the measurement error and the deviation from the initial measurement,
we conclude that the radiation damage to transparency of the samples 
is less than 1.3\% at 90\% confidence level (CL) at 9.8 Mrad dose.
\subsection{Refractive Index}
 
Samples of three refractive indices~: 1.012, 1.018 and 1.028 were 
tested. Refractive index was measured at several points,
ranging from 1 kRad to 9.8 MRad. The results are shown in 
Figure~\ref{radind}.
 
Angle $\phi$ for the aerogel samples was $90^\circ$. Typical values 
of $l$ and $d$ were 160 cm and 6 cm, respectively.
Accuracy of determining refractive index ($\Delta n$) in this method
is better than 0.001.
 
Again, we do not see any change in refractive index in any of
the samples within experimental errors \cite{problem1}.
 
\subsection{Post-irradiation Beam Test}
 
 
The three irradiated samples were later tested with the pion beam at the 
$\pi2$ beam line of the KEK-PS. Momentum of the beam was 
varied between 0.8 GeV/c and 3.5 GeV/c, from which 
refractive indices of the samples could be calculted. The
numbers perfectly agreed with the values obtained 
at the production time (the open symbols in Figure \ref{radind}).  
This gives us additional confirmation
that there has not been any radiation damage to the aerogel samples.
Using the same definition of deterioration as above, we conclude that
the radiation damage to refractive index of the samples is less than
0.001 at 90\% CL at 9.8 MRad dose. 
 
\section{Conclusion} 
 
Silica aerogels of low refractive index ($n = 1.012 \sim 1.028$)
are found to be
radiation-hard at least up to 9.8 Mrad of gamma-ray radiation. We observe
no change in transparency and refractive index within the error of
measurement after
the irradiation. Measurement accuracies were 0.8\% for
transparency and $<$0.0006 for refractive index, respectively.
90\%-CL upper limits on the radiation damage are 
1.3\% for transparency and 0.001 for refractive index, respectively.
 
 
Silica aerogels can be used in high-radiation environments, such as
$B$-factories, nuclear and heavy-ion experiments, space-station
and satellite experiments, without any fear of radiation damage.
 
\section*{Acknowledgements}
We would like to thank the BELLE Collaboration of KEK $B$-Factory
for its help in this project.
The aerogels were developed under a collaborative
research program between Matsushita Electric Works Ltd. and KEK.
One of the authors (RE) appreciates 
Drs.~S.~Kuramata (Hirosaki Univ.) and S.~Yanagida (Ibaraki Univ.)
for useful discussions on certain relevant physics issues.
We are grateful to the staff-members Dr.~F.I.~Chou, M.T.~Duo, K.W.~Fang
and Y.Y.~Wei of the Radio-isotope Division of NSTDC  
at National Tsing Hua University, Hsin-Chu (Taiwan), where the 
irradiation was conducted, for their help and co-operation. 
We are thankful to Prof.~J.C.~Peng of LANL for valuable comments
on this manuscript.
This experiment was supported in part by the grant
NSC~85-2112-M-002-034 of the Republic of China.
 

\vskip 1cm
\newpage
\section*{Table \ref{n0}, S.K.Sahu et al., NIM-A}
\begin{figure}[h!]
\epsfbox{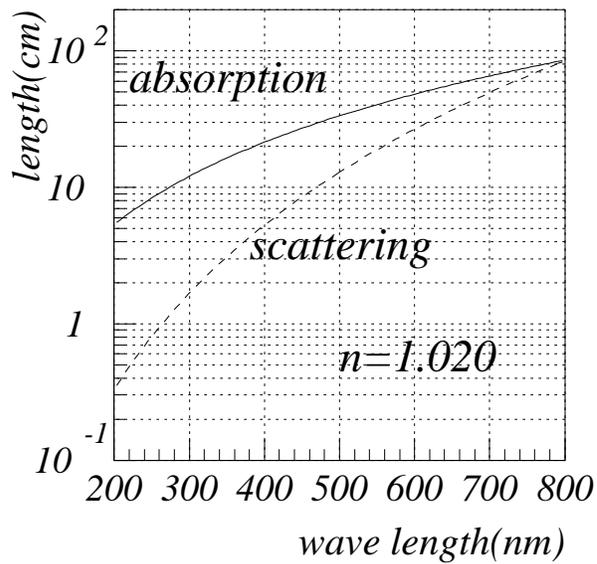}
\caption{
Absorption (solid line) and scattering (dashed line) 
spectra of a $n$ = 1.020 crystal obtained from 
fitting the transmittance curve
measured by a spectrophotometer. }
\label{transmittance}
\end{figure}

\begin{table}[h!]
\begin{tabular}{ccc}
\hline
\hline
$n$ & $N_{pe}$ & $N_0$ (cm$^{-1}$) \\
\hline
1.010 & 25.7 & 109 \\
1.015 & 37.3 & 106 \\
1.020 & 45.7 & 95 \\
\hline
\hline
\end{tabular}
\caption{
Number of photoelectrons $N_{pe}$ and 
\v Cerenkov quality factor $N_0$ for different aerogel indices, 
obtained from beam-test using $\pi^\pm$ of P = 3.5 GeV/c.}
\label{n0}
\end{table}
\newpage
\section*{Figure \ref{transmittance}, S.K.Sahu et al., NIM-A}
\begin{figure}[h!]
\epsfysize 8cm
\epsfbox{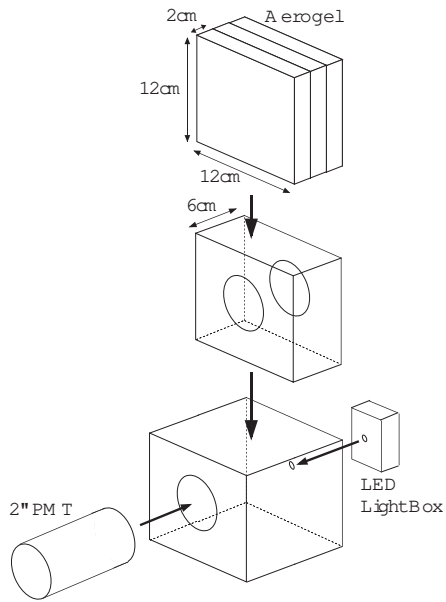}
\caption{
Experimental setup for 
measuring transmittance of aerogels.
}
\label{setupa}
\end{figure}

\newpage
\section*{Figure \ref{setupa}, S.K.Sahu et al., NIM-A}
\begin{figure}[h!]
\epsfysize 7cm
\epsfbox{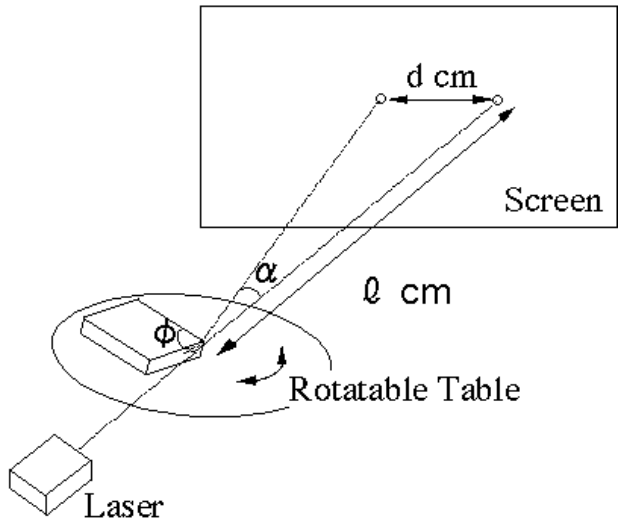}
\caption{
Experimental setup for 
measuring refractive index of aerogels.
}
\label{setupb}
\end{figure}

\newpage
\section*{Figure \ref{setupb}, S.K.Sahu et al., NIM-A}
\begin{figure}[h!]
\epsfysize 8cm
\epsfbox{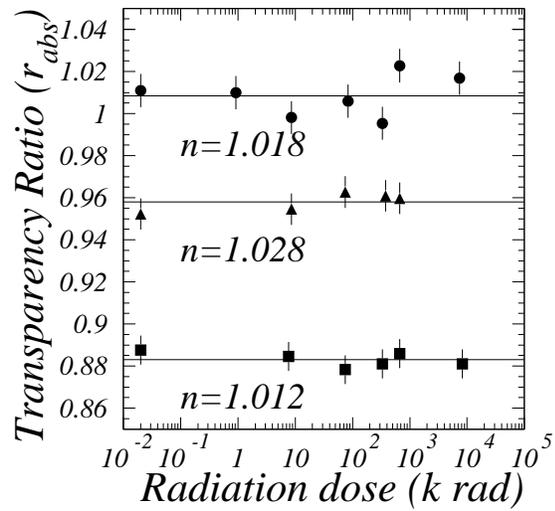}
\caption{
Transparency ratio $r_{abs}$ versus radiation dose in kilo-rad.
Each line is average value of measurements for each refractive
index sample.
The lines indicate the averages of measured values for each subsets of 
measurements.
}
\label{radtra}
\end{figure}

\newpage
\section*{Figure \ref{radtra}, S.K.Sahu et al., NIM-A}
\begin{figure}[h!]
\epsfysize 8cm
\epsfbox{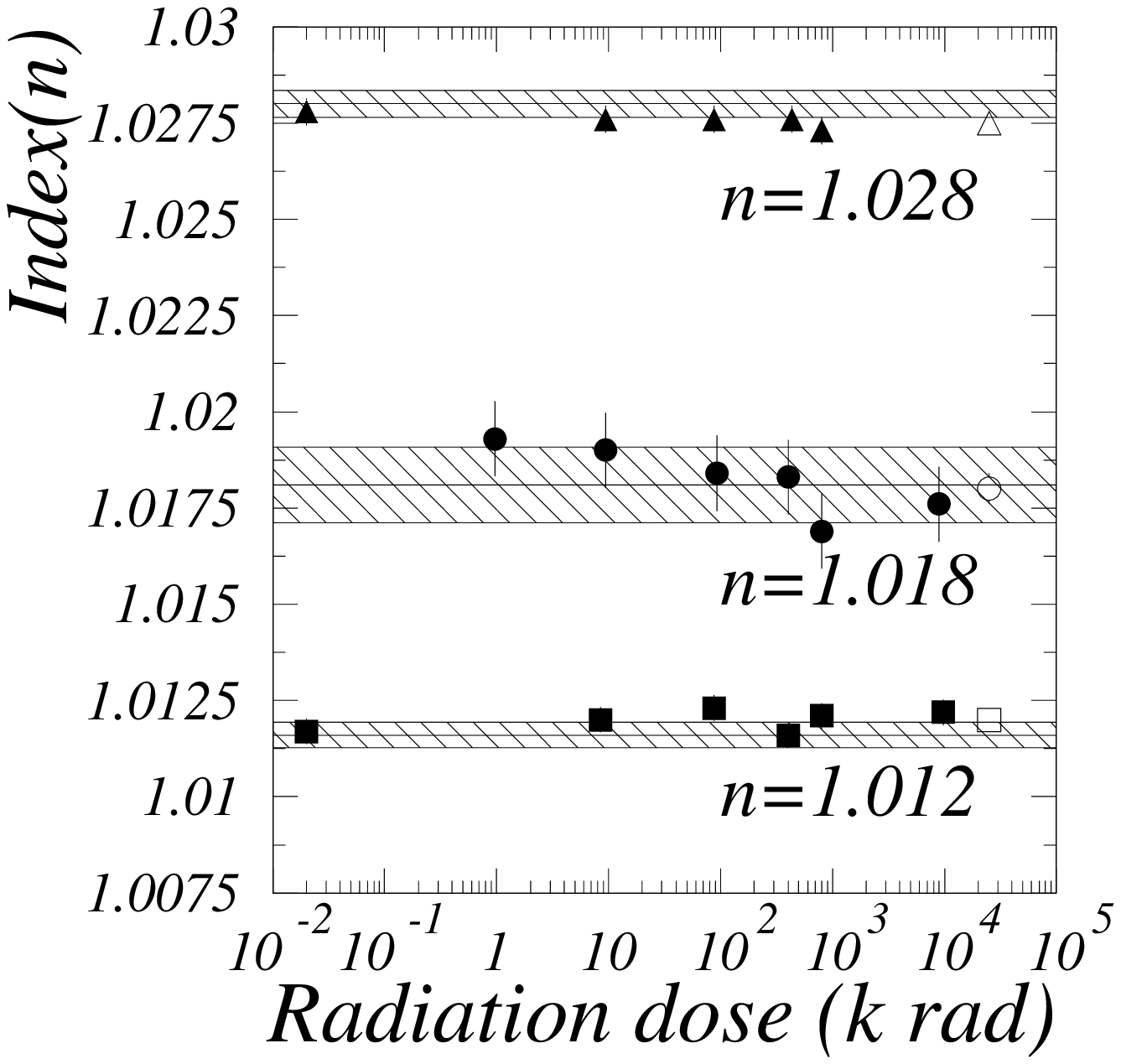}
\caption{
Refractive index versus radiation dose in kilo-rad. The solid 
line indicates the refractive index at the production time.
The shaded areas are $\pm 1 \sigma$ regions obtained from the
measurements at the production time.
The open symbols are obtained from the beam test and the closed
ones are obtained by the prism method described in the text.
}
\label{radind}
\end{figure}

\newpage
\section*{Figure \ref{radind}, S.K.Sahu et al., NIM-A}
 

\newpage
\begin{thebibliography}{99}
\bibitem{used1} D.E. Fields {\it et al.}, 
	Nucl. Instrum. Meth. {\bf A349} (1994) 431. 
\bibitem{used2} C. Lippert {\it et al.}, Nucl. Instrum. Meth. 
{\bf A333} (1993) 413.
\bibitem{used3} T. Hasegawa {\it et al.}, Nucl. Instrum. Meth. 
{\bf A342} (1994) 383.
\bibitem{used4} P. Vincent {\it et al.}, Nucl. Instrum. Meth. 
{\bf A272} (1988) 660.
\bibitem{used5} K. Maurer {\it et al.}, Nucl. Instrum. Meth. 
{\bf A224} (1984) 110.
\bibitem{used6} G. Poelz, Nucl. Instrum. Meth. {\bf A248} (1986) 118.
\bibitem{used7} P. J. Carlson {\it et al.}, Nucl. Instrum. Meth. 
{\bf A192} (1982) 209.
\bibitem{used8} H. Burkhardt {\it et al.}, Nucl. Instrum. Meth. 
{\bf A184} (1981) 319.
\bibitem{used9} C. Arnault {\it et al.}, Nucl. Instrum. Meth. 
{\bf A177} (1980) 337.
\bibitem{used10} J. P. De Brion {\it et al.}, Nucl. Instrum. Meth. 
{\bf A179} (1981) 61.
\bibitem{aerogel} I.Adachi {\it et al.}, Nucl. Instrum. Meth. 
{\bf A355}
(1995) 390.
\bibitem{slactdr} BABAR Colloboration, ``BABAR: Technical Design
Report", SLAC-R-95-457.
\bibitem{loi} Belle collaboration, ``Letter of Intent for the Belle
Collaboration", KEK Report 94-2.
\bibitem{tdr} Belle collaboration, ``Technical Design Report", 
KEK Proceedings 95-1.
\bibitem{monochrometer} Hitachi Co Ltd. model U-3210.
\bibitem{gortex} Japan Goretex Co Ltd. Spec. No. 116-027.
\bibitem{franc_tamm} ``Review of Particle Properties'',Phys. Rev. D50
(1994) 1173
\bibitem{km} M.Kobayashi and T.Maskawa, Prog. Theor. Phys. {\bf 49}
(1973) 652.
\bibitem{sanda} A.Carter and A.I.Sanda, Phys. Rev. Lett. {\bf 45} 
(1980) 952.
\bibitem{spot} T. Ypsilantis and J. Seguinot, ``RICH Outlook",
to be appeared in Proceedings of the Second Workshop on Ring Imaging
\v Cerenkov Detectors, Upsalla, Sweden, June 12-16, 1995.
\bibitem{herab} T.Lohse {\it et al.}, 
``HERA B: An experiment to study CP violation in the B system
using an internal target at the HERA proton ring: Proposal'',
DESY PRC 94-02.
\bibitem{phenix} Private communication with B.Jacak and R.S.Hayano.
\bibitem{heao} J.J.Engelmann {\it et al.}, Astron. Astrophys. {\bf 233}
(1990) 96.
\bibitem{takahashi} Private communication with T.Takahashi.
\bibitem{ting} S.C.C. Ting {\it et al.}, ``Antimatter Spectrometer in
Space'', Study of Feasibility, 1994
\bibitem{edose} We define the {\em equivalent dose} as the
amount of dose that would have been absorbed if the sample 
stopped all the $\gamma$-rays going through it. {\em Absolute dose} 
absorbed by aerogel, being
a very light material, is much smaller than the equivalent dose.
\bibitem{r329} Hamamatsu Photonics K.K., R329-05S.
\bibitem{problem1} The large fluctuation of measurements for the
n=1.020 sample is due to  its surface roughness,
which changes the effective prism angle $\alpha$.
\end{thebibliography}
\end{document}